%% file: main.tex
\documentclass[11pt,journal]{IEEEtran}

\usepackage[T1]{fontenc}
\usepackage{color}
\usepackage{subfig}
\usepackage{tabularx}
\usepackage{hyperref}
\usepackage{graphicx}

\graphicspath{{./figures/}}

\begin{document}
\input{00-title-and-authors}
\input{01-abstract}
\input{04-problem-overview}

\input{05-state-of-the-art}
\input{06-innovations-realized}
\input{04a-system}
\input{07-how-performance-measured}
\input{08-performance-results}
\input{09-implications}
\input{10a-conclusions}
\input{10-acknowledgments}

\bibliographystyle{IEEEtran}
\bibliography{references}

\end{document}

%% file: 00-title-and-authors.tex
\title{Deep Learning at 15PF: Supervised and Semi-Supervised Classification for Scientific Data}

\author{
\IEEEauthorblockN{
Thorsten Kurth\IEEEauthorrefmark{1},
Jian Zhang\IEEEauthorrefmark{2},
Nadathur Satish\IEEEauthorrefmark{3},
Ioannis Mitliagkas\IEEEauthorrefmark{2},
Evan Racah\IEEEauthorrefmark{1},
Mostofa Ali Patwary\IEEEauthorrefmark{3},
Tareq Malas\IEEEauthorrefmark{4},
Narayanan Sundaram\IEEEauthorrefmark{3},
Wahid Bhimji\IEEEauthorrefmark{1},
Mikhail Smorkalov\IEEEauthorrefmark{5},
Jack Deslippe\IEEEauthorrefmark{1},
Mikhail Shiryaev\IEEEauthorrefmark{5},
Srinivas Sridharan\IEEEauthorrefmark{6},
Prabhat\IEEEauthorrefmark{1},
Pradeep Dubey\IEEEauthorrefmark{3}
}

\thanks{\IEEEauthorblockA{\IEEEauthorrefmark{1}Lawrence Berkeley National Laboratory, 1 Cyclotron Road, M/S 59R4010A, Berkeley, CA 94720}
\IEEEauthorblockA{\IEEEauthorrefmark{2}Stanford University, Gates Computer Science, 353 Serra Mall, Stanford, CA 94305}
\IEEEauthorblockA{\IEEEauthorrefmark{3}Intel Corporation, 2200 Mission College Blvd., Santa Clara, CA 95054}
\IEEEauthorblockA{\IEEEauthorrefmark{4}Intel Corporation, 2111 NE 25th Avenue, JF-1-06, Hillsboro, OR 97124}
\IEEEauthorblockA{\IEEEauthorrefmark{5}Intel Corporation, Turgeneva Str. 30, Nizhny Novgorod, Russian Federation 603024}
\IEEEauthorblockA{\IEEEauthorrefmark{6}Intel Corporation, 136 Airport Road, Bangalore, Karnataka, India 560007}
}
}

\maketitle

%% file: 01-abstract.tex
\begin{abstract}
This paper presents the first, 15-PetaFLOP Deep Learning system for solving scientific pattern classification problems on contemporary HPC architectures. We develop supervised convolutional architectures for discriminating signals in high-energy physics data as well as semi-supervised architectures for localizing and classifying extreme weather in climate data. Our Intelcaffe-based implementation obtains $\sim$2TFLOP/s on a single Cori Phase-II Xeon-Phi node. We use a hybrid strategy employing synchronous node-groups, while using asynchronous communication across groups. We use this strategy to scale training of a single model to $\sim$9600 Xeon-Phi nodes; obtaining peak performance of 11.73-15.07 PFLOP/s and sustained performance of 11.41-13.27 PFLOP/s.  At scale, our HEP architecture produces state-of-the-art classification accuracy on a dataset with 10M images, exceeding that achieved by selections on high-level physics-motivated features. Our semi-supervised architecture successfully extracts weather patterns in a 15TB climate dataset. Our results demonstrate that Deep Learning can be optimized and scaled effectively on many-core, HPC systems. 
\end{abstract}

%% file: 04-problem-overview.tex
\section{Deep Learning for Science}
In recent years, Deep Learning (DL) has enabled fundamental breakthroughs in computer vision, speech recognition and control system problems, thereby enabling a number of novel commercial applications. At their core, these applications solve classification and regression problems, tasks which are shared by numerous scientific domains. For example, problems in identifying galaxies, screening medical images, predicting cosmological constants, material properties and protein structure prediction all involve learning a complex hierarchy of features, and predicting a class label, or regressing a numerical quantity. 
We assert that that Deep Learning is poised to have a major impact on domain sciences, but there are  unique challenges that need to be overcome first.

The primary challenge is in analyzing massive quantities of complex, multi-variate scientific data. Current Deep Learning implementations can take days to converge on O(10) GB datasets; contemporary scientific datasets are TBs-PBs in size. Scientific datasets often contain dozens of channels/variables, which is in contrast to the small number of channels in images or audio data. Scientists need to be able to leverage parallel computational resources to get reasonable turnaround times for training Deep Neural Networks (DNNs). It is therefore imperative that DL software delivers good performance not only on a single node but is also scalable across a large number of nodes. We now elaborate on two scientific drivers that motivate our optimization and scaling efforts. 

\subsection{Supervised Learning for HEP}
\label{sec:hep}
A major aim of experimental high-energy physics (HEP) is to find rare signals of new particles produced at accelerators such as the Large Hadron Collider (LHC) at CERN, where protons are accelerated to high-energies and collided together to produce resulting particles within highly-instrumented detectors, such as the ATLAS and CMS experiments. Improvements in classifying these collisions could aid discoveries that would overturn our understanding of the universe at the most fundamental level. 
Neural Networks have been used in HEP for some time \cite{HepNN,HepNN2}. Recently attention has focused on deep learning
to tackle the increase in detector resolutions and data rates. Particles produced by LHC collisions (occurring every 25ns) propagate, decay and deposit energy in different detector parts, so creating signals in 100s of millions of channels, with each collision forming an independent `event'.
Data from the surface of the cylindrical detector can be represented as a sparse 2D image, with data from different layers of instrumentation as channels in that image. We use the energy deposited in the ``electromagnetic'', and ``hadronic calorimeters'', and the number of ``tracks'' formed from the ``inner detector'' in that region as three channels. This is similar to the approach of  \cite{jetimage1}\cite{jetimage2} except that we use large images covering the entire detector, and use these directly for classifying entire events rather than individual objects.  

The HEP community have simulations of the underlying physics processes and the detector response that can be used for training networks.  For this paper, we generate events to match those used for a particular analysis searching for new massive supersymmetric particles in multi-jet final states at the LHC \cite{ATLAS:2016nij}. We use the Pythia event generator \cite {Pythia} interfaced to the Delphes fast detector simulation \cite{Delphes} (with fast jet \cite{fastjet}) to generate events for two classes, corresponding to the new-physics `signal' (6.4M events) and the most prevalent known-physics `background' (64M events). Before training our network we apply some of the physics selections of \cite{ATLAS:2016nij} to filter images to those more challenging to discriminate, resulting in a training sample of around 10M events. We compare the performance of our deep network to our own implementation of the selections of \cite{ATLAS:2016nij} as a baseline benchmark. We have verified that the samples and baseline selections give performance comparable to that in \cite{ATLAS:2016nij} providing a meaningful benchmark even though those selections were not tuned for these datasets. 

\subsection{Semi-Supervised Learning for Climate}

Climate change is one of the most important challenges facing humanity in the 21st century;  climate simulations provide a unique approach for understanding the future impact of various carbon emission scenarios and intervention strategies. Modern Climate simulation codes produce massive datasets: a single 30-year run from the CAM5 25-km resolution model produces 100TBs of multi-variate data\cite{Wehner:2005}. In this paper, we are interested in the task of finding extreme weather events in such large datasets. Providing an objective, quantitative tool for finding extreme weather patterns will help climate scientists in understanding trends in such weather patterns in the future (i.e. Do we expect more Category 4/5 hurricanes to make landfall in the 21st century?), and conduct detection and attribution studies (i.e. Is the chance in Tropical Cyclone activity attributable to anthropogenic emissions, as opposed to being an intrinsic property of the climate system?). 

The field of climate science typically relies on heuristics, and expert-specified multi-variate threshold conditions for specifying extremes \cite{knutson2010,Lavers,imilast}. We formulate this task as that of pattern classification, and employ Deep Learning based methods. 
The problem can be formulated as that of object recognition in images, the difference being that climate images have 16 or more 'channels', and their underlying statistics are quite different from natural images. Consequently, we cannot leverage pre-trained weights from contemporary networks such as VGG or AlexNet. Earlier work conducted by \cite{yunjie:abda16} demonstrates that convolutional architectures can solve the pattern classification task for cropped, centered image patches. In this work we develop a \emph{unified, semi-supervised} architecture for handling all extreme weather patterns and develop a methodology for predicting bounding boxes. Most importantly, our method provides an opportunity to discover new weather patterns that might have few/no labeled examples. 

\begin{table}
\centering
  \begin{tabular}{|l|c|c|c|c|}
    \hline
     & pixels & channels & \#images & Volume \\ \hline
    HEP & 228x228 & 3 & 10M & 7.4TB \\ \hline
    Climate & 768x768 & 16 & 0.4M & 15TB \\ \hline
  \end{tabular}
  \caption{Characteristics of datasets used.}
  \label{tab:datasets}
\end{table}

This paper makes the following contributions:
\begin{itemize}
\item{We develop Deep Learning models which not only solve the problem at hand to desired precision but are also scalable to a large number of nodes. This includes for example to not use layers with large dense weights such as batch normalization or fully connected units.}
\item{We develop highly optimized Deep Learning software that can process complex scientific datasets on the Intel Xeon Phi architecture}
\item{We build a system based on a hybrid asynchronous approach to scale Deep Learning to the full scale of the Cori supercomputer ($\sim$9600 Xeon Phi nodes)}
\item{We demonstrate supervised classification on a 7.4 TB High-Energy Physics dataset}
\item{We develop a novel, semi-supervised architecture, and apply it to detect and learn new patterns on a 15 TB climate dataset}
\item{We obtain a peak performance of 11.73-15.07 PFLOP/s and sustained performance of 11.41-13.27 PFLOP/s for our two problems}
\end{itemize}
While our exploration is conducted in the context of two concrete applications, we believe that our approach, and the resulting lessons learned, can be generalized to a much broader class of data analytics problems in science.

%% file: 05-state-of-the-art.tex
\section{Current State of the Art}
From a HPC perspective, we can look at deep learning from two dimensions: first, how efficiently can deep learning be mapped to a single compute node; and second, how it scales across a cluster of compute nodes.

\subsection{Deep Learning on single node}
The core computation in deep learning algorithms is dominated by dense linear algebra in the form of matrix multiply and convolution operations. While well-optimized libraries such as implementations of BLAS and LaPACK have long existed for use in HPC applications, the shapes and sizes of the operands differ significantly for deep learning. Hence specific libraries with support for tall-skinny matrix multiplies and convolutions with multiple small filters have been developed for various architectures such as NVIDIA GPUs \cite{CUDNNv5} and CPU architectures \cite{MKL2017DeepLearning, libxsmm}. 

The hardware efficiency of these kernels 
heavily depends on input data sizes and  model parameters (weight matrix dimensions, number of convolutions, convolution strides, padding, etc). DeepBench \cite{DeepBench} is a recently developed benchmark from Baidu that captures best known performance of deep learning kernels with varied input sizes and model parameters on NVIDIA GPUs and Intel$\textsuperscript{\textregistered}$ Xeon Phi$\textsuperscript{\texttrademark}${\footnote{Intel, Xeon and Intel Xeon Phi are trademarks of Intel Corporation in the U.S. and/or other countries.}}. Their results show that while performance can be as high as 75-80\% of peak flops for some kernels, decreasing minibatch size (dimension 'N' for matrix multiply and convolutions) results in significant efficiency drops to as low as 20-30\% (at minibatch sizes of 4-16) on all architectures. As we shall see, this has implications on performance at scale.

\subsection{Deep Learning on multiple nodes}

There have been many attempts to scale deep learning models across a cluster of nodes~\cite{baidudeepspeech2, Dean2012, Firecaffe, Dipankar16, CNTK17}. In this work, we focus on scaling the training of a \emph{single model} across a cluster as opposed to the embarassingly parallel problem of training independent models \cite{ORNL}. 
We discuss two common architectures, shown in Figure~\ref{fig:sync_async}.

\subsubsection{Synchronous-parallel architectures}
Synchronous systems use synchronization barriers and force computational nodes to perform every update step in lock-step (See Figure \ref{fig:sync_async}).
Typically, data parallelism is used where different nodes split a big mini-batch of samples, each processing a chunk of the data. 
Recent papers that have attempted to scale synchronous deep learning have stopped at a few hundred nodes \cite{Dipankar16, Firecaffe,mxnet_slides}, with the scalability depending on the computation to communication ratio, the speed of the hardware and the quality of the interconnect. Aside from communication
there are other factors that limit synchronous scaling:

\paragraph{Batch size}
Most systems use some variant of SGD with batch sizes that range from $64$ to $1024$. 
Large batch sizes have been shown to cause slowdown in convergence~\cite{hadjis2016omnivore}, and degrade the generalization properties of the trained model~\cite{keskar2016large}.
The batch size is a limit on the number of nodes in data-parallel synchronous systems. 

\paragraph{Stragglers}
Since a synchronization barrier is used, the duration of the iteration depends on the slowest node. Variability in the computation needed per sample, OS jitter and, importantly, variations in the throughput and latency in the interconnect leads to significant load imbalance. This effect gets worse with scale. 

\subsubsection{Asynchronous and hybrid architectures}
\label{sec:soa_asynchrony}
\begin{figure}[!t]
\centering
\includegraphics[width=3.3in]{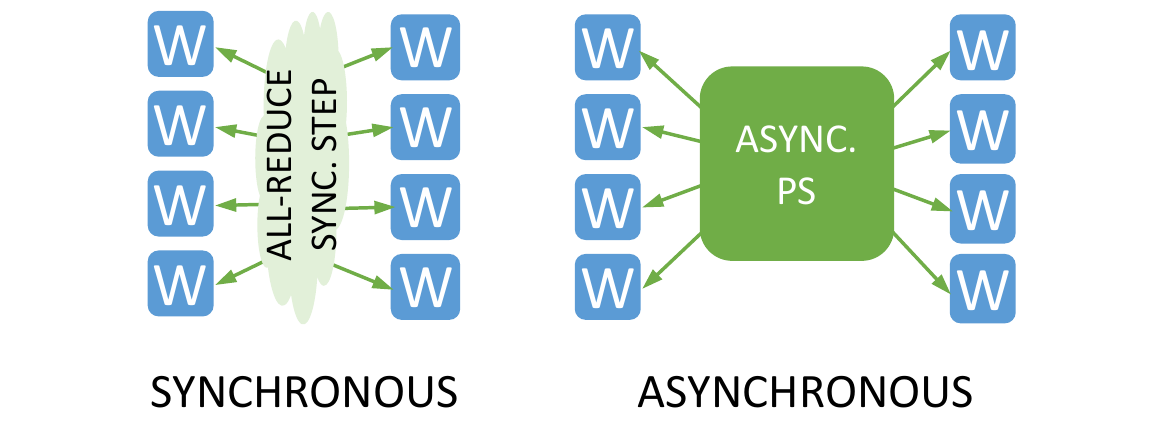}
\caption{Example architectures.}
\label{fig:sync_async}
\end{figure}
Conceptually, asynchronous architectures \cite{tsitsiklis1986distributed,recht2011hogwild} remove the synchronization barriers. 
Each node works on its own iteration (mini-batch) and produces independent updates to the model.
Those updates are sent to a central parameter store, the {\em parameter server} (PS), illustrated in Figure~\ref{fig:sync_async}.
The PS applies the updates to the model in the order they are received, and sends back the updated model to the worker where the update originated.
Asynchronous systems do not suffer from straggler effects and are not limited by the total batch size in the same way that synchronous systems are, an important property at scale. 
Asynchronous methods are known to give significant computational benefits in large-scale systems \cite{dean2012large,chilimbi2014project}.
Recent work \cite{mitliagkas2016asynchrony} sheds new light on the convergence properties of such systems and shows the importance of momentum tuning for convergence.

\paragraph{Performance tradeoff}
The main side-effect of asynchrony is the use of out-of-date gradients: 
each update is computed based on an older version of the model and then sent to the PS to be applied on the latest model.
The number of updates that other workers perform between the time a worker reads the model and the time it sends its own update to the PS is called {\em staleness}.
Asynchronous systems may need more iterations to solution, due to staleness: we say they have worse {\em statistical efficiency}  \cite{DBLP:journals/pvldb/ZhangR14,hadjis2016omnivore}.
Synchronous systems typically take longer per iteration due to the straggler effect: they have worse {\em hardware efficiency}.

\paragraph{Hybrid architectures}
The trade-off between statistical efficiency vs. hardware efficiency suggests a third kind of architecture: a {\em hybrid} system \cite{hadjis2016omnivore}.
In this architecture, worker nodes coalesce into separate {\em compute groups}. 
Each compute group 
follows a synchronous architecture: the workers split a mini-batch among themselves and produce a single update to the model.
There is no synchronization across compute groups.
A parameter server (PS) holds the model and each compute group communicates its updates to the PS asynchronously.
Given a cluster of fixed size, the number of compute groups (and their size) is a knob that controls the amount of asynchrony in the system.
We can tune the amount of asynchrony along with the other hyper-parameters to find the optimal configuration.
We use this hybrid architecture in our paper, as described in 
Section~\ref{sec:our_hybrid_architecture}.

%% file: 06-innovations-realized.tex
\section{Innovations}

\begin{table*}[th]
\centering
  \begin{tabular}{|l|c|c|c|c|}
    \hline
    Architecture & Input & Layer details  & Output & Parameters size\\ \hline
    Supervised HEP & 224x224x3 & 5xconv-pool,1xfully-connected & class probability & 2.3MiB \\ \hline
    Semi-supervised Climate & 768x768x16 & 9xconv,5xDeconv &coordinates, class, confidence & 302.1 MiB\\ \hline
  \end{tabular}
  \caption{Specification of DNN architectures used in this study.}
  \label{tab:networks}
\end{table*}

\subsection{HEP architecture}
We formulate the HEP problem as a binary image classification task. We use a Convolutional Neural Net comprised of 5 convolution+pooling units with rectified linear unit (ReLU) activation functions \cite{hahnloser2000digital,he2015delving}. The kernel sizes used in the convolutional layers are 3x3 pixels with strides 1x1 and 128 filters per layer. In the pooling layers we use 2x2 kernels with strides 2x2. We use max pooling in the first four layers and use global average pooling in the last convolutional layer. 
The output of the global pooling layer is fed into a single fully connected layer which projects the resulting 128-dimensional vector into a two-dimensional vector on which a softmax function is applied to determine the class probabilities for signal and background. We use softmax with cross-entropy as the loss function. 
We further employ the ADAM optimizer\cite{kingma2014adam} as the solver. ADAM requires less parameter tuning than Stochastic Gradient Descent and suppresses high norm variability between gradients of different layers by adaptively adjusting the learning rate. 

\subsection{Climate architecture}
We formulate the climate problem as semi-supervised bounding box regression adapted from \cite{racah2016semi}, which is inspired by \cite{redmon2016you, liu2016ssd, ren2015faster}. Essentially, we have a fully supervised convolutional network for bounding box regression and an unsupervised convolutional autoencoder. These two networks share various layers, so the extra unlabelled data input to the autoencoder can help improve the bounding box regression task. We use a a series of strided convolutions to learn coarse, downsampled features of the input climate simulations. We call this series of convolutions the encoder of the network. At every location in the features, we compute 4 scores (confidence, class, x and y position of bottom left corner of box, and height and width of box) using a convolution layer for each score. At inference time we keep only the boxes corresponding to confidences greater than 0.8. For the unsupervised part of our architecture, we use the same encoder layers, but use the coarse features as input to a series of deconvolutional layers, which we call the decoder. The decoder attempts to reconstruct the input climate image from the coarse features. The objective function attempts to simultaneously minimize the confidence of areas without a box, maximize those with a box, maximize the the probability of the correct class for areas with a box, minimize the scale and location offset of the predicted box to the real box and minimize the reconstruction error of the autoencoder. As a solver, we use stochastic gradient descent with momentum.

\subsection{Single-node performance on manycore architectures}
In this work, we used the Intel distribution of Caffe \cite{intelcaffe} to train our models. This distribution links in the  Intel MKL 2017 library~\cite{MKL2017DeepLearning} with optimized deep learning primitives for Intel Xeon Phi. For our semi-supervised climate network, we needed optimized implementations of deconvolution that were not available. We used the fact that the convolutions in the backward pass can be used to compute the deconvolutions of the forward pass and vice-versa in order to develop optimized deconvolution implementations. These layers perform very similarly to the corresponding convolution layers.

\begin{figure}[!t]
\centering
\includegraphics[width=3.3in]{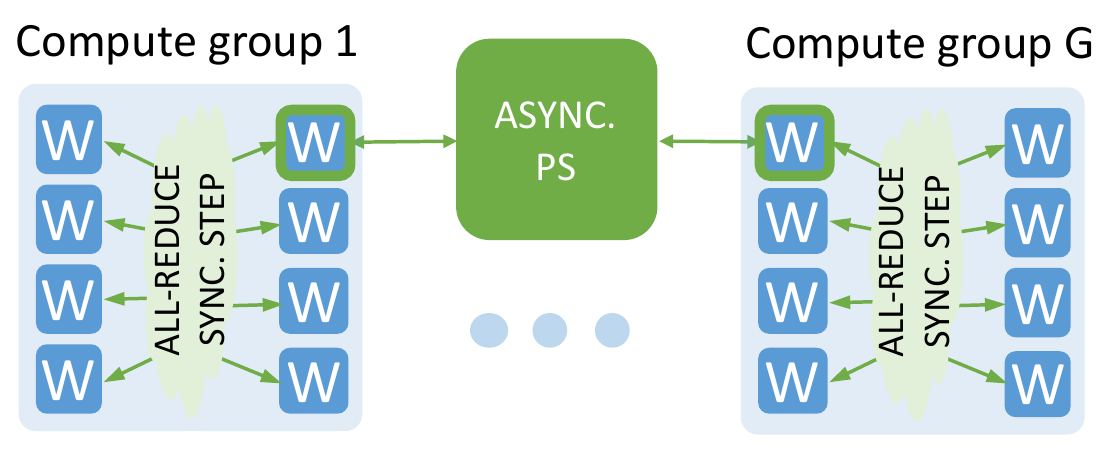}
\caption{Hybrid architecture example.}
\label{fig:hybrid}
\end{figure}

\subsection{Multi-node scaling with synchronous approach}
We utilize the new Intel$\textsuperscript{\textregistered}$~Machine Learning Scalability Library (MLSL) \cite {mlsl} for our multi-node implementation. This handles all communication required to perform training in a synchronous setting, and enables different forms of parallelism - both data and model parallelism - to be applied to different layers of the network without the user/developer worrying about communication details. In this work, we deal with either fully convolutional networks or those with very small fully connected layers, so we only use data parallelism which is well suited for such layers. MLSL also introduces performance improvements over vanilla MPI implementations using endpoints - proxy threads/processes which drive communication on behalf of the MPI rank and enable better utilization of network bandwidth. 
Results with this library have not been reported at large scales of more than a few hundred nodes; in this work we attempt to scale this out to thousands of nodes.

\subsection{Multi-node scaling with hybrid approach}
\label{sec:our_hybrid_architecture}
\begin{figure}[!t]
\centering
\includegraphics[width=3.3in]{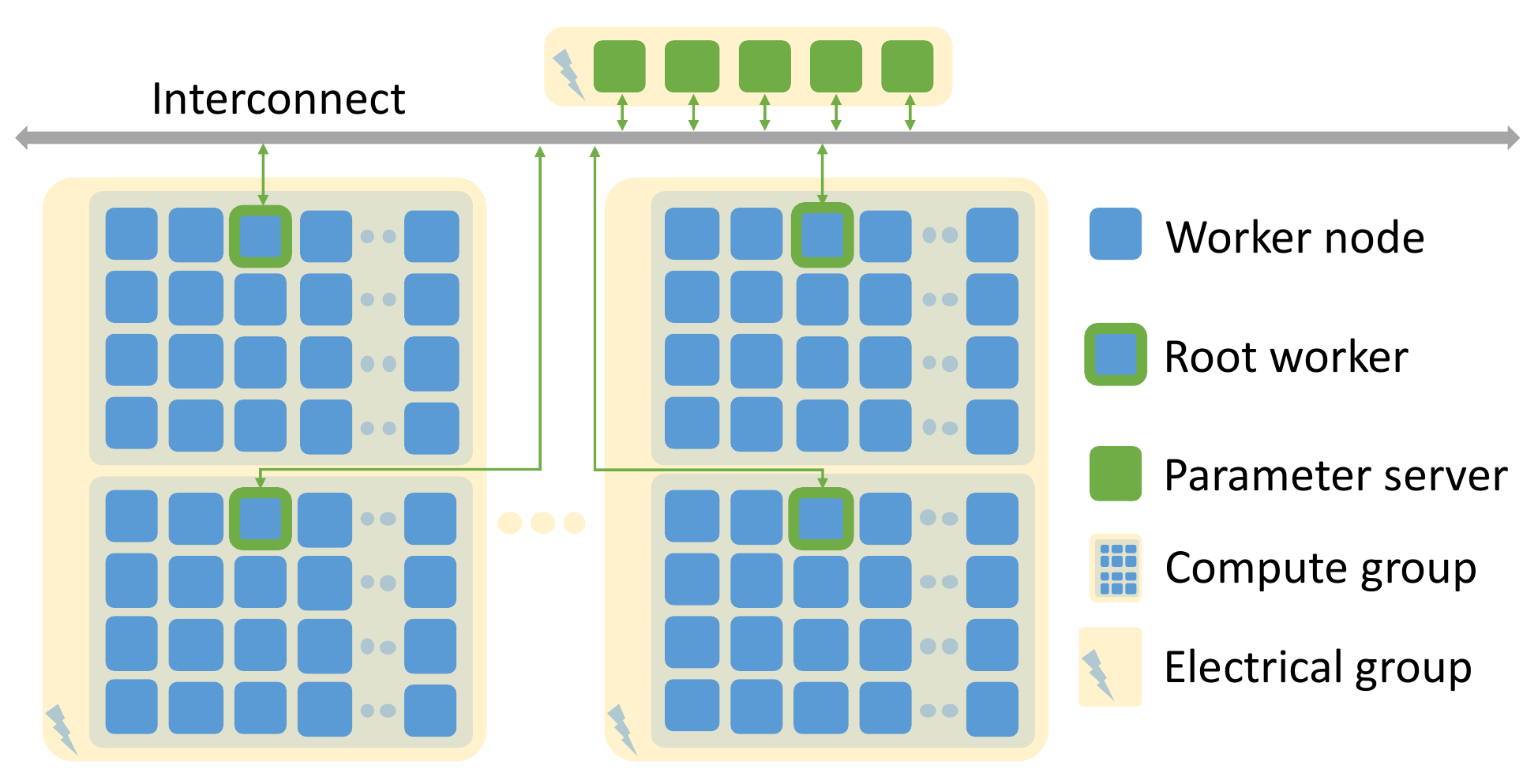}
\caption{Topological placement on Cori Phase II.}
\label{fig:placement}
\end{figure}
In Section~\ref{sec:soa_asynchrony} we outlined the limitations of fully synchronous systems that motivate asynchronous architectures.
Asynchronous systems are not limited by the total batch size in the same way that synchronous systems are. 
Furthermore, asynchrony provides an added layer of resilience to node failures and the straggler effect.
In this section we describe the hybrid architecture we use in our system and discuss some of its novel elements.

Our architecture is inspired by recently proposed hybrid approaches \cite{hadjis2016omnivore}, depicted in Figure~\ref{fig:hybrid}.
Nodes are organized into compute groups.
Parallelization is synchronous within (using all-reduce), but asynchronous across groups via a set of parameter servers.
The number and size of compute groups, is a knob which controls the level of asynchrony, and allows us to tune asynchrony and momentum jointly, as per recent theoretical guidelines \cite{mitliagkas2016asynchrony}.
Figure~\ref{fig:placement} shows an ideal placement of nodes and compute groups on Cori.\footnote{For simplicity PSs are shown in their own electrical group, however this is not typically the case.}
All-reduce operations are used to get the aggregate model update from all workers in the group.
Then a single node per group, called the {\em root node} is responsible for communicating the update to the parameter servers, receiving the new model, and broadcasting it back to the group.

\paragraph{Extreme Scale}
Our work is the first instance of a hybrid architecture that scales to thousands of nodes. Previous implementations were designed (and typically deployed) on dozens or hundreds of commodity machines. For the present work, we deployed our implementation on configurations of up to 9600 nodes on an HPC system.

\paragraph{Use of MLSL library}
MLSL does not natively support asynchronous communication. Specifically, all nodes are assumed to communicate with each other and the default library did not allow us to dedicate some subset of nodes for parameter servers. In this work, we extended MLSL to enable our hybrid implementation. Specifically, we extended MLSL to facilitate node placement into disjoint communication groups and dedicating nodes as parameter servers. Our new MLSL primitives allow for efficient overlaying of group communication and endpoint communication with the parameter server.

\begin{figure}[!t]
\centering
\includegraphics[width=3.3in]{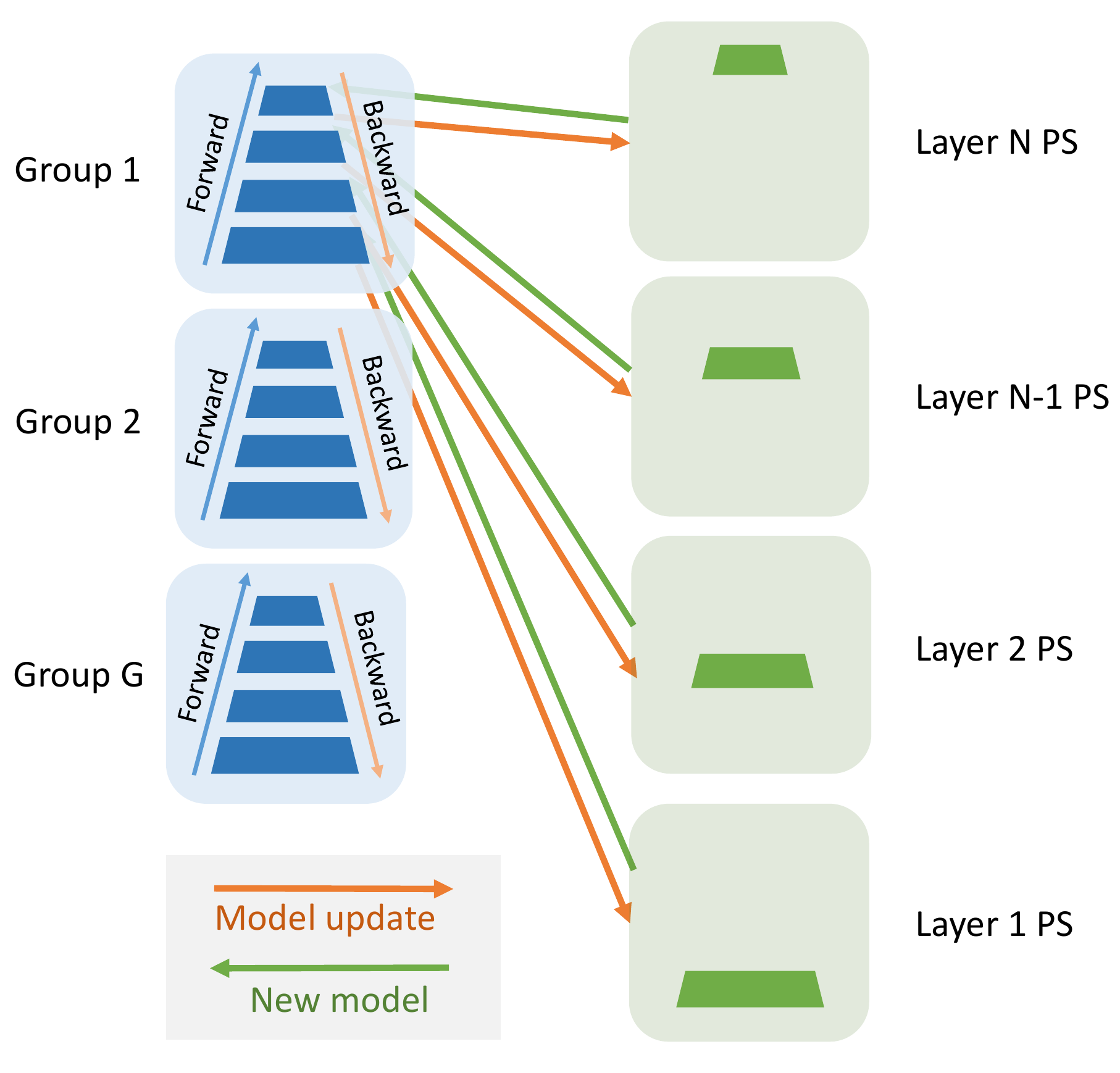}
\caption{We assign a dedicated parameter server to each trainable layer of the network. Each group exchanges data with the PS for the corresponding layer. For clarity, we only depict the communication patterns for Group 1.}
\label{fig:workers_ps_layers}
\end{figure}

\paragraph{Dedicated parameter servers for each layer}
The parameter server needs to be able to handle the volume of network traffic and computation for the updates originating from multiple compute groups and for very large models.
To reduce the chances of PS saturation, we dedicate a parameter server to each trainable layer in the network (Figure~\ref{fig:workers_ps_layers}).
We can consider each compute group as a bigger, more powerful node, that performs the usual forward and backward pass operations on the layers of the network. 
The backward pass generates a gradient (model update) for each layer of the network.
That update is communicated to its {\em dedicated parameter server}, the update is performed and the model communicated back to the same compute group.

%% file: 04a-system.tex
\section{Cori Phase II}
All experiments reported in this study are conducted on the Cori Phase II system at NERSC. Cori is a Cray XC40 supercomputer comprised of 9,688 self-hosted Intel Xeon Phi\textsuperscript{\texttrademark} 7250 (Knight's Landing, KNL) compute nodes.
Each KNL processor includes 68 cores running at 1.4GHz and capable of hosting 4 HyperThreads for a total of 272 threads per node. 

The peak performance for single precision can be computed as: (9688 KNLs) x (68 Cores) x (1.4 GHz Clock Speed) x (64 FLOPs / Cycle) = 59 PetaFLOP/s. However, for sustained AVX work, the clock-speed drops to 1.2 GHz, yielding a sustained peak performance of: 50.6 PetaFLOP/s.

Each out-of-order superscalar core has a private 32KiB L1 cache and two 512-bit wide vector processing units (supporting the AVX-512 instruction set\footnote{This includes the subsets F, CD, ER, PF but not VL, BW, DQ, IFMA, VBMI.}).  Each pair of cores (a ``tile'') shares a 1MiB L2 cache and  each node has 96GiB of DDR4 memory and 16GiB of on-package high bandwidth (MCDRAM) memory.
The MCDRAM memory can be configured into different modes, where the most interesting being \textit{cache} mode in which the MCDRAM acts as a 16GiB L3 cache on DRAM.
Additionally, MCDRAM can be configured in \textit{flat} mode in which the user can address the MCDRAM as a second NUMA node.
The on-chip directory can be configured into a number of modes, but in this publication we only consider \textit{quad} mode, i.e. in {\it quad-cache}, all cores are in a single NUMA domain with MCDRAM acting as a cache on DDR4 main memory. Furthermore, Cori features the Cray Aries low-latency, high-bandwidth interconnect utilizing the dragonfly topology.

%% file: 07-how-performance-measured.tex
\section{Performance Measurement}
\label{sec:measurement}
We count the executed FLOPs using Intel$\textsuperscript{\textregistered}$ Software Development Emulator (SDE)~\cite{sde}.
SDE distinguishes the precision of the FLOP operations and the actual executed FLOPs in the masked SIMD instructions of the code.
We use SDE to count the executed single-precision flops in the computational kernels (i.e, the neural network layers) of a single node.
Given that all the nodes execute these layers the same number of times and using the same problem size, we compute the total FLOPs by multiplying the single node FLOPs by the number of nodes. The counted FLOPs constitute the vast majority of the application's FLOP operations.
The application time is spent in an iterative training loop, where the computation performed in each training iteration is the same. However, in some iterations, a checkpointing is performed to save the current trained model to the filesystem; this imposes some overhead on runtime.
We measure the wall clock time per iteration to obtain the flop rate (i.e. iteration's measured FLOPS / iteration's time). The peak flop rate is obtained from the fastest iteration, while the sustained flop rate is computed from the best average iteration time in a contiguous window of iterations. 

In the following section, we present the results of training the HEP and climate networks on the Intel Xeon Phi nodes of the Cori supercomputer. All our experiments use 66 of the 68 cores on each node, with 2 being reserved for the OS. All our experiments deal with single precision data and model parameters.

%% file: 08-performance-results.tex
\section{Performance Results}

\subsection{Single node performance}

\begin{figure*}[!t]
\centering
\subfloat[HEP]{\includegraphics[width=3.3in]{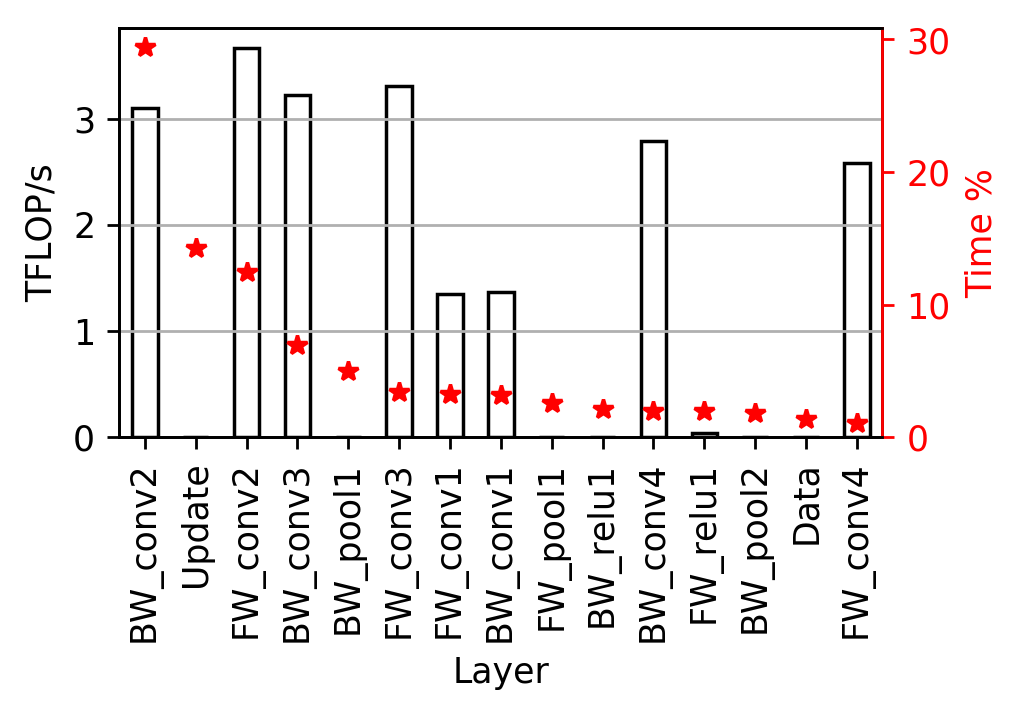}%
\label{fig:hep_single_b8}}
\hfil
\subfloat[Climate]{\includegraphics[width=3.3in]{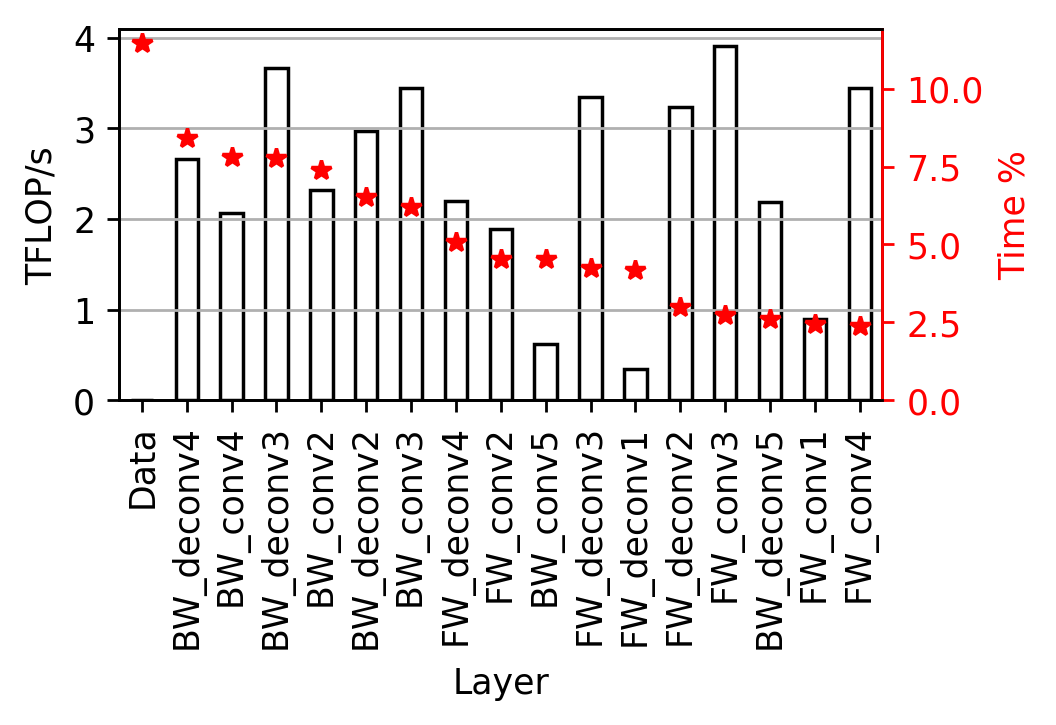}
\label{fig:climate_single_b8}}
\caption{Single node runtime and flop rate of the top time consuming components, with batch size 8}
\label{fig:single_node}
\end{figure*}

Figures~\ref{fig:hep_single_b8} and~\ref{fig:climate_single_b8} show the flop rates and time spent in various layers for HEP and Climate networks. For a batch size of 8 images, the overall flop rate of the HEP network stands at 1.90 TFLOP/s, while that of the Climate network stands at 2.09 TFLOP/s. For both networks, most of the runtime is spent in convolutional layers, which can obtain between 3.5 TFLOP/s for layers with many channels, and around 1.25 TFLOP/s on the initial layers with very few channels. As mentioned previously in DeepBench \cite{DeepBench}, the shapes of the parameters and inputs to a layer can affect performance significantly; we observe that in our experiments.

For the HEP network, about 12.5\% of the runtime is spent in the solver update routine which applies the update to the weights and adjusts hyper-parameters for the next iteration. This step spends time in operations like copying models to keep history that do not contribute to flops. The overhead of this step is insignificant ($<$ 2\%) in the climate network. For the climate network, time spent in I/O (13\%) for loading the data is significant; recall that climate problem consists of high resolution, 16-channel data. In comparison, the I/O time is much lower (~$2$ \%) for the HEP network, which has low resolution, 3-channel data. We have identified two bottlenecks in our current I/O configuration: first, I/O throughput from a single Xeon Phi core is relatively slow, second, the current HDF5 library is not multi-threaded. We will address these limitations in future work. 

\subsection{Multi-node scaling}
We now report on scaling experiments conducts on Cori Phase II. 

\subsubsection{Strong Scaling}
\begin{figure*}[!t]
\centering
\subfloat[HEP]{\includegraphics[width=3.3in]{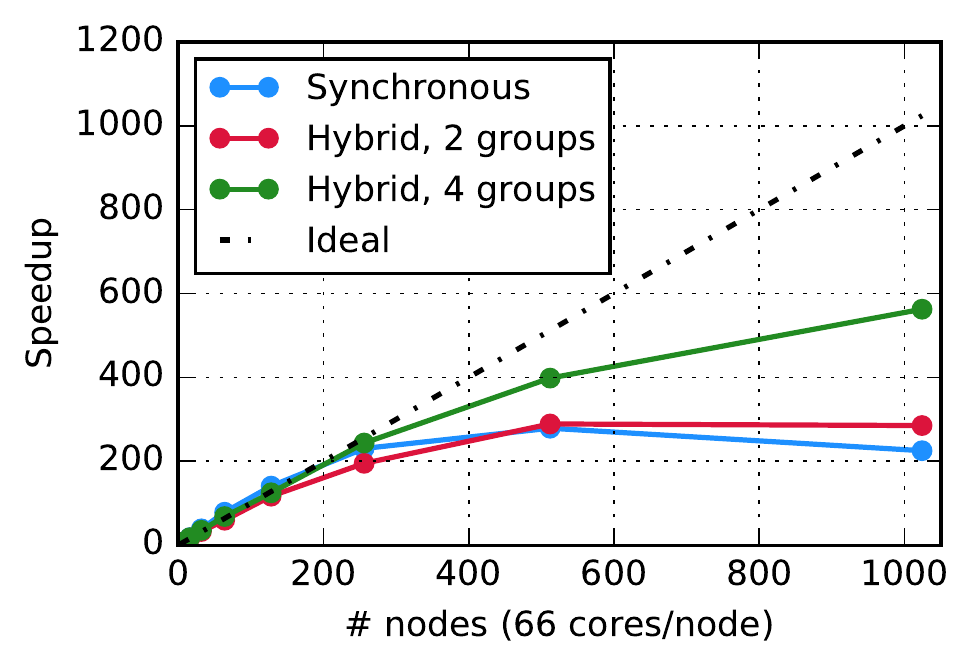}%
\label{fig:hep_strong_scaling}}
\hfil
\subfloat[Climate]{\includegraphics[width=3.3in]{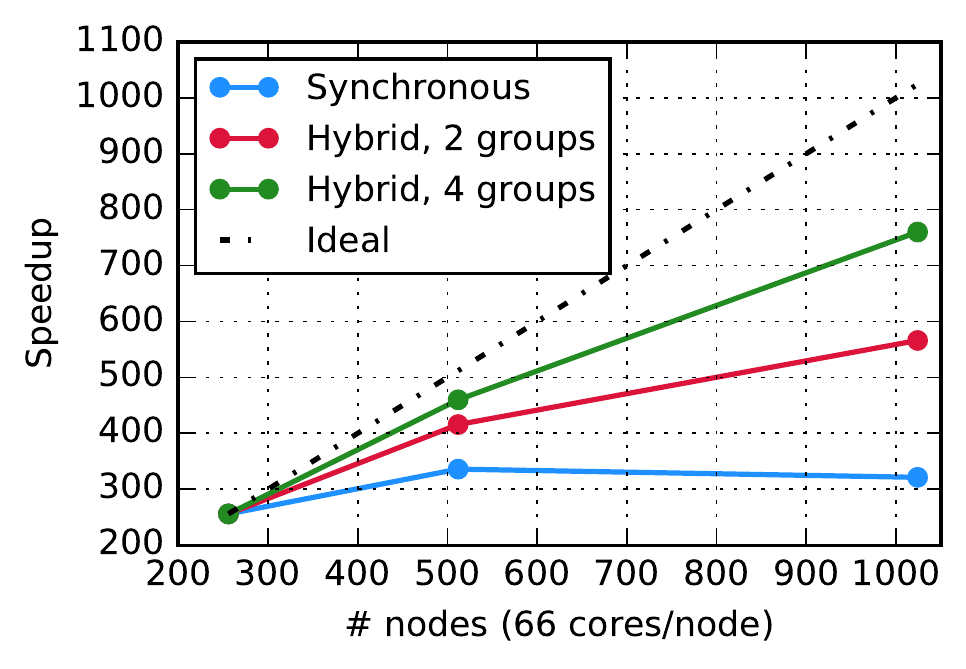}%
\label{fig:climate_strong_scaling}}
\caption{Strong scaling results for synchronous and hybrid approaches (batch size = 2048 per synchronous group).}
\label{fig:strong_scaling}
\end{figure*}

The strong scaling configuration (involving keeping the overall batch size per update step fixed while varying the number of nodes) is a natural use-case for deep learning. Figure~\ref{fig:strong_scaling} shows the strong scaling results for HEP and climate networks. We show 3 configurations:
1 synchronous group, 2 and 4 hybrid groups; and show scalability from 1 to 1024 nodes. 
We use a batch size of 2048 per update. For the synchronous configuration, all nodes split the batch of 2048 images; for hybrid configurations, each compute group independently updates the model and is assigned a complete batch.
Figure~\ref{fig:hep_strong_scaling} shows that the synchronous algorithm does not scale past 256 nodes -- 1024 node performance is somewhat worse than for 256. The scalability improves moderately for 2 hybrid groups, which saturates at 280x beyond 512 nodes, and more significantly with 4 hybrid groups, with about 580x scaling at 1024 nodes.
We observe similar trends for the climate network in Figure~\ref{fig:climate_strong_scaling} - the synchronous algorithm scales only to a maximum of 320x at 512 nodes and stops scaling beyond that point. The 2 and 4 group hybrid groups continue scaling to 1024 nodes; with scalability improving from 580x (on 1024 nodes) for 2 hybrid groups to 780x for 4 hybrid groups. 
There are two main reasons for this: one, in hybrid algorithms, only a subset of nodes need to synchronize at each time step; this reduces communication costs and straggler effects. Second, the minibatch size per node is higher for the hybrid approaches resulting in better single node performance. Scaling for our hybrid approaches is still not linear due to the single node performance drop from reduced minibatch sizes at scale.

\subsubsection{Weak Scaling}

\begin{figure*}[!t]
\centering
\subfloat[HEP]{\includegraphics[width=3.3in]{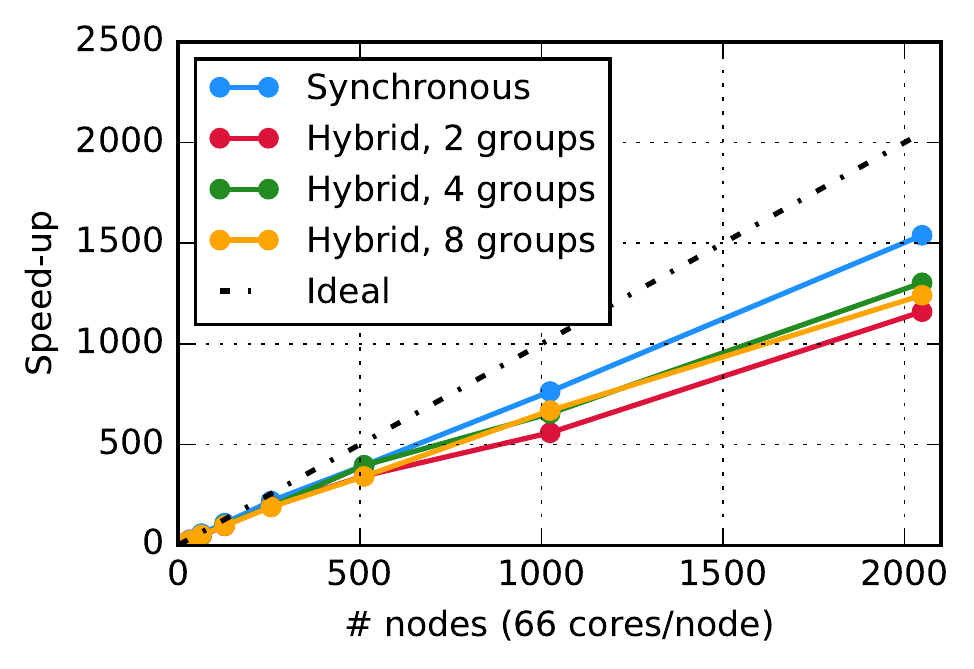}%
\label{fig:hep_weak_scaling}}
\hfil
\subfloat[Climate]{\includegraphics[width=3.3in]{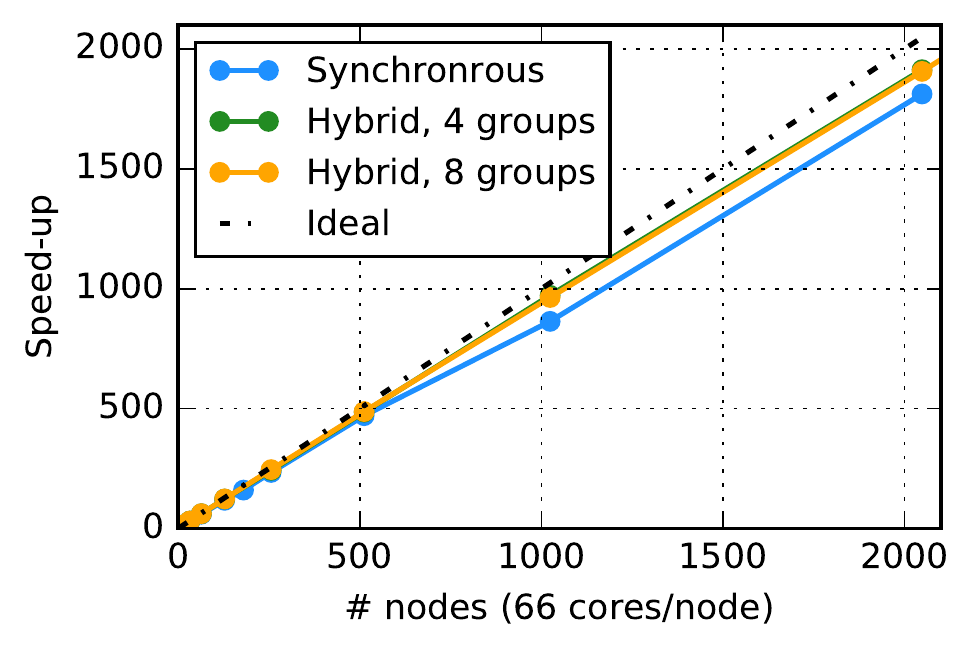}%
\label{fig:climate_weak_scaling}}
\caption{Weak scaling results for synchronous and hybrid approaches (batch size = 8 per node).}
\label{fig:weak_scaling}
\vspace*{-0.2in}
\end{figure*}

Figure~\ref{fig:hep_weak_scaling} shows weak scaling for the HEP network, where we keep a constant batch size (8 per node) across all configurations (synchronous and hybrid). On scaling from 1 to 2048 nodes, we find that the performance scales sub-linearly for all configurations: about 575-750x speed-up on 1024 nodes; and about 1150-1250x speed-up on 2048 nodes for asynchronous configurations. We note that the synchronous speed-up on 2048 nodes stands at about 1500x. In contrast, the weak scaling results for the climate network in Figure~\ref{fig:climate_weak_scaling} are near-linear (1750x for synchronous and about 1850x for hybrid configurations). Our analysis indicates significant variability in runtime across iterations for HEP at scale, leading to sublinear scaling. An average convolution layer in HEP takes about 12 ms to execute; at the end of which nodes need to synchronize and reduce a small model of $\sim$590 KB. Even a small jitter in communication times can lead to significant variability in this scenario. Hybrid approaches, where we have two additional communication steps (to and from the PS) are more affected by this variability, leading to reduced scaling. Our climate model 
takes on average over 300 ms per convolution layer, leading to less frequent communication and impact from jitter - we observe slightly better scaling for hybrid over synchronous configurations due to reduced straggler effects. 

\subsubsection{Overall Performance}
For the HEP network, we obtained a {\bf peak} throughput (as described in Section~\ref{sec:measurement}) of {\bf 11.73~PFLOP/s} for a configuration of 9600 total nodes (9594 compute nodes plus 6 parameter servers) split into 9 groups, with each group using a minibatch of 1066. This corresponds to a speedup of 6173x over single node performance. The {\bf sustained} throughput as measured over a 100 iteration timespan is {\bf 11.41~PFLOP/s}. This corresponds to an average per-iteration runtime of about 106 ms for processing a minibatch.

For the climate network, we obtained a {\bf peak} throughput of {\bf 15.07~PFLOP/s} for a configuration of 9622 total nodes (9608 compute nodes plus 14 parameter servers) split into 8 groups, with each group using a minibatch of 9608. This corresponds to a speedup of 7205X over single node performance. The {\bf sustained} throughput as measured over a 10 iteration span is about {\bf 13.27~PFLOP/s}, corresponding to a speedup of an average per-iteration runtime of 12.16 seconds. The sustained throughput computed includes the overhead of storing a model snapshot to disk once in 10 iterations, causing slowdowns.

\subsubsection{Time to Train}
\label{subsubsec:time_to_train}
\begin{figure}[!t]
\centering
\includegraphics[width=3.3in]{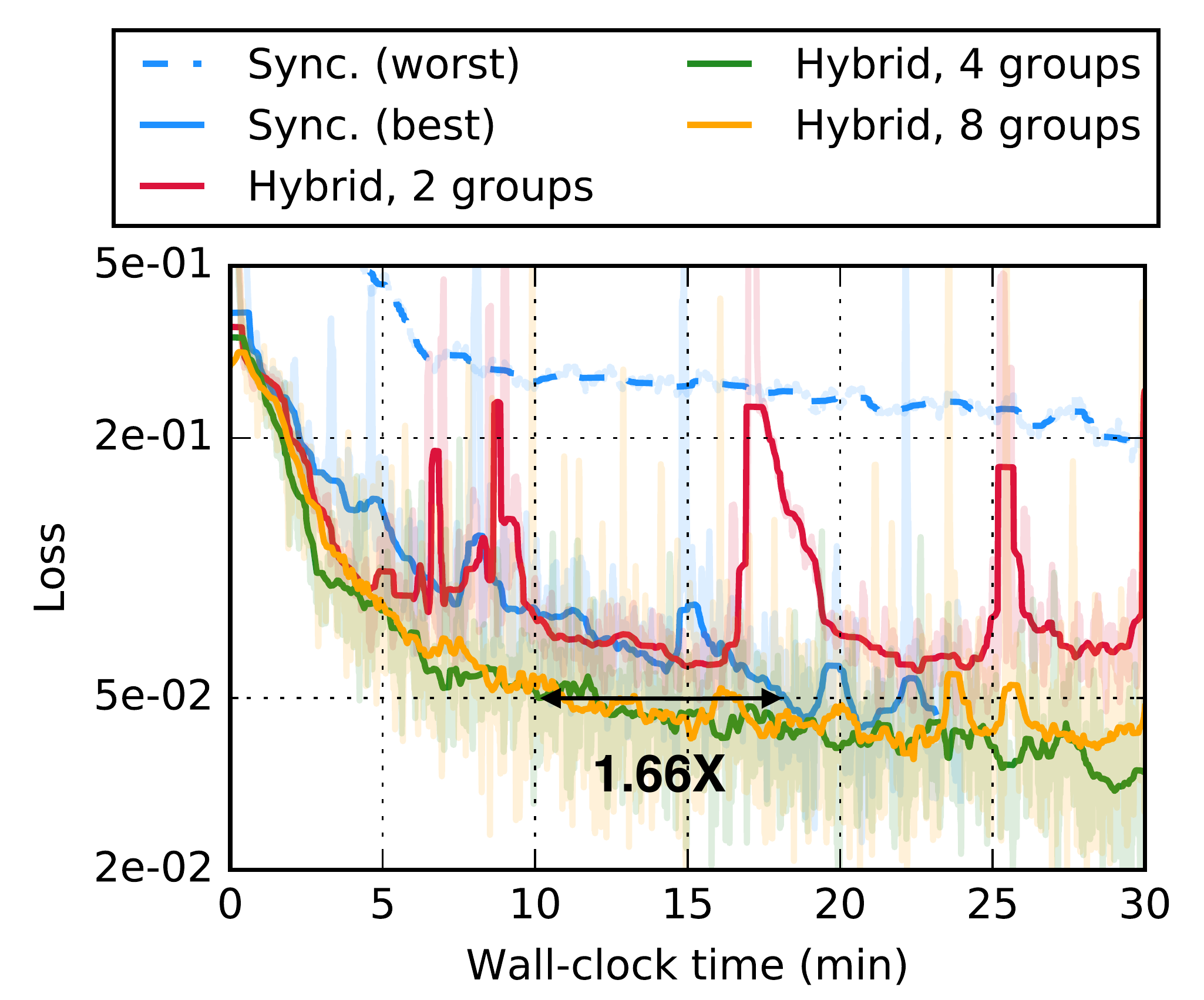}
\caption{Training losses vs wall clock time for HEP on 1K nodes. Comparing synchronous configuration to 2,4 and 8 groups.}
\label{fig:hep_wall_clock}
\end{figure}
Figure~\ref{fig:hep_wall_clock} reports the result of different training runs on the HEP network using $1024$ worker nodes.
We fix the total batch to $1024$ and try a fully synchronous run, and three hybrid runs with $2,4,8$ groups. We use the Adam update and tune its learning rate in the following range: $[1e-4, 1e-3]$. 
For the synchronous setting we fix its momentum to $0.9$, but for hybrid runs we tune the momentum on a discrete set of values (0.0, 0.4, 0.7) to account for the momentum contributed by asynchrony \cite{mitliagkas2016asynchrony}.
We report the measured training loss over wall-clock for the best configurations.
For the synchronous setting, we report (for the same best hyper-parameter configuration) the best and worst run out of $3$.
We report wall-clock time speedups with respect to a loss of $0.05$ that beats the baseline for HEP (as defined in Section~\ref{sec:hep}).
We establish that the {best hybrid configuration achieves the target loss in about 10 minutes}, which is about  {$1.66\times$ faster than the best sync run}. 
The worst sync run is many times slower.
We attribute this, as well as some of the jumps observed in the loss curves of the $2$-group case to variability in individual node performance when running on $1$K nodes. Note that without additional hyperparameter tuning, we achieve a speedup of 11x in time to convergence for going from 64 to 1024 nodes, which is in line with expectations from weak scaling (cf. Figure~\ref{fig:hep_weak_scaling}).

\section{Science Results}

\subsection{HEP Science Result}
For the HEP classification problem, it is important to achieve a high signal efficiency at a very low acceptance of the much more prevalent background class. Our benchmark analysis, which is based on selections on high-level physics-derived features, achieves a true-positive rate of 42\% at a false-positive rate of 0.02\%. To evaluate our results we compare the true-positive rate at this same very low false-positive rate. For the hybrid configuration described in section \ref{subsubsec:time_to_train}, we achieve a rate of 72\% which represents a \textbf{1.7x improvement} over our benchmark. 
For the full-system runs reported here, even with reduced runtime and without extensive tuning for accuracy, the SGD solver outperforms our benchmark by 1.3X. The capability to achieve high sensitivities to new-physics signals from classification on low-level detector quantities, without the need to design, reconstruct, or tune, high-level features offers considerable potential for enabling new-physics discoveries in future HEP analyses.

\subsection{Climate Science Result}
\begin{figure}[!t]
\centering
\includegraphics[width=3.2in]{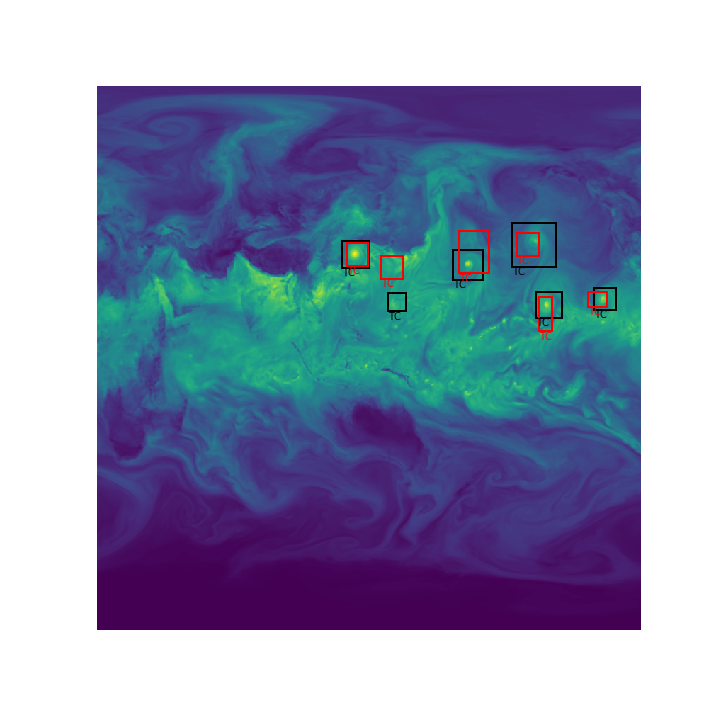}
\caption{Results from  plotting the network's most confident (>95\%) box predictions on an image for integrated water vapor (TMQ) from the test set for the climate problem. Black bounding boxes show ground truth; Red boxes are predictions by the network.}
\label{fig:climateresult}
\end{figure}

Figure \ref{fig:climateresult} presents a sample image that illustrates the ability of our semi-supervised architecture to produce bounding boxes and class labels. In the figure, the architecture does a good job of localizing and identifying tropical cyclones. We are working on generating additional metrics for assessing the accuracy of bounding boxes for known classes (including extra-tropical cyclones and atmospheric rivers). More importantly, we are evaluating the ability of the architecture to discover \emph{novel} weather patterns. Since this is fundamentally new approach for pattern detection in the climate science community, we do not have a well-established benchmark to compare our results to.

%% file: 09-implications.tex
\section{Implications}

\subsection{Deep Learning on HPC}
\label{sec:dl_hpc}

To the best of our knowledge, our work is the first successful attempt at scaling Deep Learning on large, many-core HPC systems. We share a number of insights from this unique exercise.

First, at a scale of thousands of nodes, we found significant variability in runtimes across runs, which could be as high as 30\%. The probability of one of the thousands of nodes failing or degrading during the run is non-zero. In this work, we report runs where we did not encounter complete node failures. We note that even a single node failure can cause complete failure of synchronous runs; hybrid runs are much more resilient since only one of the compute groups gets affected. However, even in hybrid runs, if model updates from one of the compute groups lags significantly behind others, it can result in "jumps" in the overall loss and accuracy that we have highlighted in Figure~\ref{fig:hep_wall_clock}.

Second, current architectures and software stacks for deep learning are still not as mature as the traditional HPC application stack. Specifically, performance on small batch sizes (essential for scale out) has not been completely optimized in many frameworks. 
Further, the state of the art in deep learning kernel implementations is rapidly evolving with new algorithms like Winograd~\cite{winograd} and FFT based algorithms. We did not experiment with such algorithms in this work; studying the impact on per-node performance and scale out behaviour of these algorithms is a direction for future research. 

There has been a lot of discussion surrounding training with quantized weights and activations~\cite{hubara16,courbariaux14}. The statistical implications of low precision training are still being explored~\cite{gupta15,gysel16}, with various forms of stochastic rounding being of critical importance in convergence. While supercomputers with architectures supporting low precision computations in hardware are not yet present, we believe that such systems have the potential to further accelerate training time for our applications.

\subsection{Deep Learning for Science}
We believe that science domains that can readily generate vast amounts of representative training data (via simulators) stand to benefit immediately from progress in DL methods. In other scientific domains, unsupervised, and semi-supervised learning are key challenges for the future. 
In both cases, it is unreasonable to expect scientists to be conversant in the art of hyper-parameter tuning. Hybrid schemes, like the one presented in this paper, add an extra parameter to be tuned, which stresses the need for principled momentum tuning approaches, an active area of research (eg.\cite{hadjis2016omnivore} and recently \cite{zhang2017yellowfin}). With hyper-parameter tuning taken care of, higher-level libraries such as Spearmint \cite{snoek2012} can be used for automating the search for network architectures.\\
We also note that more aggressive optimizations involving computing in low-precision and communicating high-order bits of weight updates are poorly understood with regards to their implications for classification and regression accuracy for \emph{scientific} datasets. A similar story holds with regards to deployment of DL models. Unlike commercial applications where a sparse/compact representation of the model needs to be deployed in-situ, scientific applications will typically utilize DL models within the context of the HPC/Datacenter environment. Nevertheless, the field of Deep Learning is evolving rapidly, and we look forward to adopting advances in the near future.

%% file: 10a-conclusions.tex
\section{Conclusions}
This paper has presented the first 15-PetaFLOP Deep Learning software running on HPC platforms. We have utilized IntelCaffe to obtain $\sim$2 TF on single Xeon Phi nodes. We utilize a hybrid strategy employing synchronous groups, and asynchronous communication among them to scale the training of a single model to $\sim$9600 Cori Phase II nodes. We apply this framework to solve real-world supervised and semi-supervised patterns classification problems in HEP and Climate Science. Our work demonstrates that manycore HPC platforms can be successfully used to accelerate Deep Learning, opening the gateway for broader adoption by the domain science community. Our results are not limited to the specific applications mentioned in this paper, but they extend to other kinds of models such as ResNets~\cite{he2016deep} and LSTM~\cite{lstm_orig,lstm_imp}, although the optimal configuration between synchronous and asynchronous is expected to be model dependent. This highlights the importance of a flexible, hybrid architecture in achieving the best performance for a diverse set of problems.

%% file: 10-acknowledgments.tex
\section*{Acknowledgments}

This research used resources of the National Energy Research Scientific Computing Center (NERSC). This manuscript has been authored by an author at Lawrence Berkeley National Laboratory under Contract No. DE-AC02-05CH11231 with the U.S. Department of Energy. The U.S. Government retains, and the publisher, by accepting the article for publication, acknowledges, that the U.S. Government retains a non-exclusive, paid-up, irrevocable, world-wide license to publish or reproduce the published form of this manuscript, or allow others to do so, for U.S. Government purposes.
We would like to thank Doug Jacobsen, Brandon Cook, Tina Declerck, David Paul and Rebecca Hartman-Baker for assisting with and troubleshooting Cori reservations. 
We would like to acknowledge Christopher Beckham, Tegan Maharaj and Christopher Pal, Yunjie Liu and Michael Wehner for help with preparing the climate architecture and dataset. We would like to acknowledge Steve Farrell for assistance with preparing HEP datasets and Ben Nachman and Brian Amadio for physics input on those datasets. Christopher R\'e's group at Stanford was the source of valuable advice on asynchrony. We would like to thank Srinivas Sridharan, Mikhail Smorkalov, Mikhail Shiryaev and Dipankar Das for their help in integrating and modifying Intel MLSL.